\documentclass{Proceeding}

\usepackage[utf8]{inputenc} 
\usepackage{amsmath}

\providecommand{\keywords}[1]{\textit{Keywords:} #1}

\begin{document}
\vspace*{0cm}
\title{Detailed background dynamics and trans-planckian effects in loop quantum cosmology}

\author{Killian Martineau}

\address{Laboratoire de Physique Subatomique et de Cosmologie, Universit\'e Grenoble-Alpes,\\
Grenoble, 38000, France\\
E-mail: killian.martineau@lpsc.in2p3.fr\\
http://lpsc.in2p3.fr/index.php/en/}

\maketitle\abstracts{
Cosmology appears as the most promising way to test and constrain quantum gravity theories. Loop quantum gravity is among the most advanced attempts to perform a non-perturbative quantization of general relativity. Its cosmological counterpart, loop quantum cosmology, has clear predictions both for the cosmological background and for the perturbations. In particular, the initial Big Bang singularity is replaced by a bounce due to quantum geometry effects. In this proceeding I will focus on new results obtained in loop quantum cosmology: i) the prediction of the duration of inflation as a function of all the unknown parameters of the model and ii) new primordial power spectra obtained with modified dispersion relations accounting for trans-planckian effects.}


\vspace*{0.35cm}

\section{Introduction}

Loop quantum gravity \cite{lqg1,lqg2} (LQG) might be the most advanced non-perturbative and background-invariant quantization of general relativity.  Loop quantum cosmology \cite{lqc1} (LQC) is a quantum theory inspired by the LQG quantization scheme that takes into account the cosmological symmetries. 
Its background dynamics,  governed by the effective modified Friedmann equation

\begin{equation}
H^2=\frac{\kappa}{3}\rho\left(1-\frac{\rho}{\rho_c}\right)~,
\label{Friedmann LQC}
\end{equation}

leads to a bounce and is now well established. This prediction has been confirmed by numerical simulations \cite{Diener}, but also by calculations in group field theory \cite{Oriti:2016qtz}, underlining the robustness of the model.
In this equation $H$ is the Hubble parameter, $\kappa= 8 \pi$, $\rho$ stands for the energy density, and $\rho_c\sim \rho_{Pl}$ for its value at the bounce.

The situation is however less clear for perturbations. Not only because different settings are being considered (mainly the {\it deformed algebra} \cite{Bojowald:2011aa} and the {\it dressed metric} \cite{Agullo1} approaches) but most importantly because trans-planckian effects are often neglected or not treated as a dominant process. It seems that too much emphasis has been put on {\it density} effects and not enough on {\it length} effects. Even if the former do trigger the bounce the latter cannot be ignored in a theory that modifies the space structure at the Planck length.
In this proceeding I will therefore focus on two new results in LQC : i) the prediction of the duration of inflation as a function of all the unknown parameters of the model and ii) new primordial power spectra obtained with modified dispersion relations accounting for trans-planckian effects.
 
\section{Exhaustive investigation of the inflation duration in the loop quantum cosmological framework}

Remarkably, inflation is a strong attractor and appears naturally in the LQC model as long as the matter content of the universe is assumed to be a massive scalar field. One of the most interesting consequence of this cosmological framework is therefore that the inflation duration can, to some extent, be predicted. 

However, even at the background level, three main uncertainties remain to be addressed systematically. 
The first one is related with the way to set initial conditions. There are mainly two schools of thought : the first one sets them in the remote past of the contracting branch whereas the other one sets them at the bounce. Since those initial conditions are in a one-to-one correspondence with each other the key point is to look for a variable to which a  known probability distribution function (PDF) can be assigned. 
The second uncertainty is associated with the amount of anisotropic shear at the bounce, which is expected to play an important role in any bouncing model.  Finally the third main uncertainty is the inflaton potential shape, as the matter content of the universe is not predicted by LQG (but can be partially experimentally determined). 

Within this framework we have performed an exhaustive investigation of the duration of inflation by varying those three unknowns \cite{Martineau:2017sti}.

When dealing with the initial conditions issue, it is in our opinion more appealing to set them in the remote past of the classical contracting branch of the Universe. In this case, a flat PDF can easily be associated to the initial phase of the scalar field.
If we do so, and if the inflaton potential is confining, the duration of inflation appears to be severely constrained and, most interestingly, the predicted number of e-folds is not much higher than the minimum value required by observations. This result is an important feature of the loop quantum cosmological model and underlines its strong predictive power for a massive scalar field. This predictive power is even increased when anisotropies are taken into account as the number of e-folds corresponding to the PDF mean value decreases. 
However the LQC predictive power is basically lost if initial conditions are set at the bounce as there is apparently no variable to which a known PDF can be assigned.

In summary, once the inflaton potential is determined and if initial conditions are set in the contracting branch, in agreement with the intuition of causality,  there is an obviously interesting predictive power of LQC for the inflation duration. This predictive power is stronger when the potential is confining and weaker if it features a plateau-like shape. The initial amount of shear remains however unknown but this is not necessarily a problem if the inflaton potential is confining as the predicted \textit{e}-folds number is then restricted to a small interval, bounded from below by observations and from above by the isotropic case.

\section{A first step towards the inflationary trans-planckian problem treatment in loop quantum cosmology}

The trans-planckian problem is a well-known cosmological issue \cite{Martin:2000xs} that cannot be ignored in the LQG framework. As soon as the number of e-folds of inflation is higher than 70, all modes of physical interest were highly trans-planckien at the bounce time.

As a first elementary step to account for trans-planckian effects, we have suggested the use of modified dispersion relations (MDRs) in the LQC framework \cite{Martineau:2017tdx}. This is obviously not the final word on this question and trans-planckian effects should ideally be considered in the full theory setting but we believe that MDRs are a meaningful first step.\\

\begin{figure}[h]
\begin{center}
\includegraphics[scale=0.47]{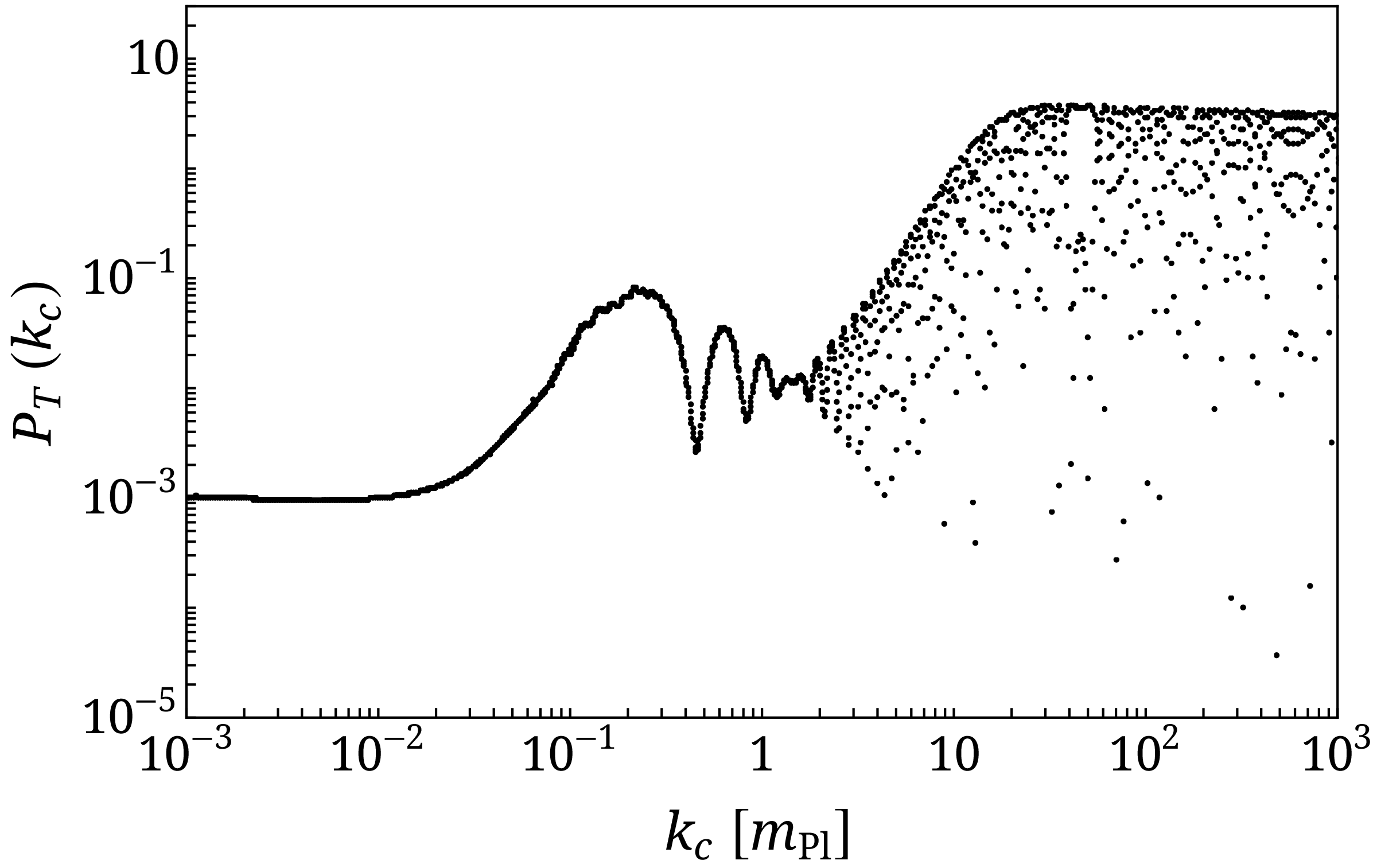}
\end{center}
\caption{Primordial tensor power spectrum in the deformed algebra approach with the Unruh-like MDR $~~$ $\mathcal{F}(k_{\varphi}) = k_{0} \tanh\left[\left(k_{c}/(a(t)k_{0})\right)^{p}\right]^{1/p}$, $k_{0}=10$ and $p=1$.}
\label{tensor power spectrum}
\end{figure}

\begin{figure}[h]
\begin{center}
\includegraphics[scale=0.218]{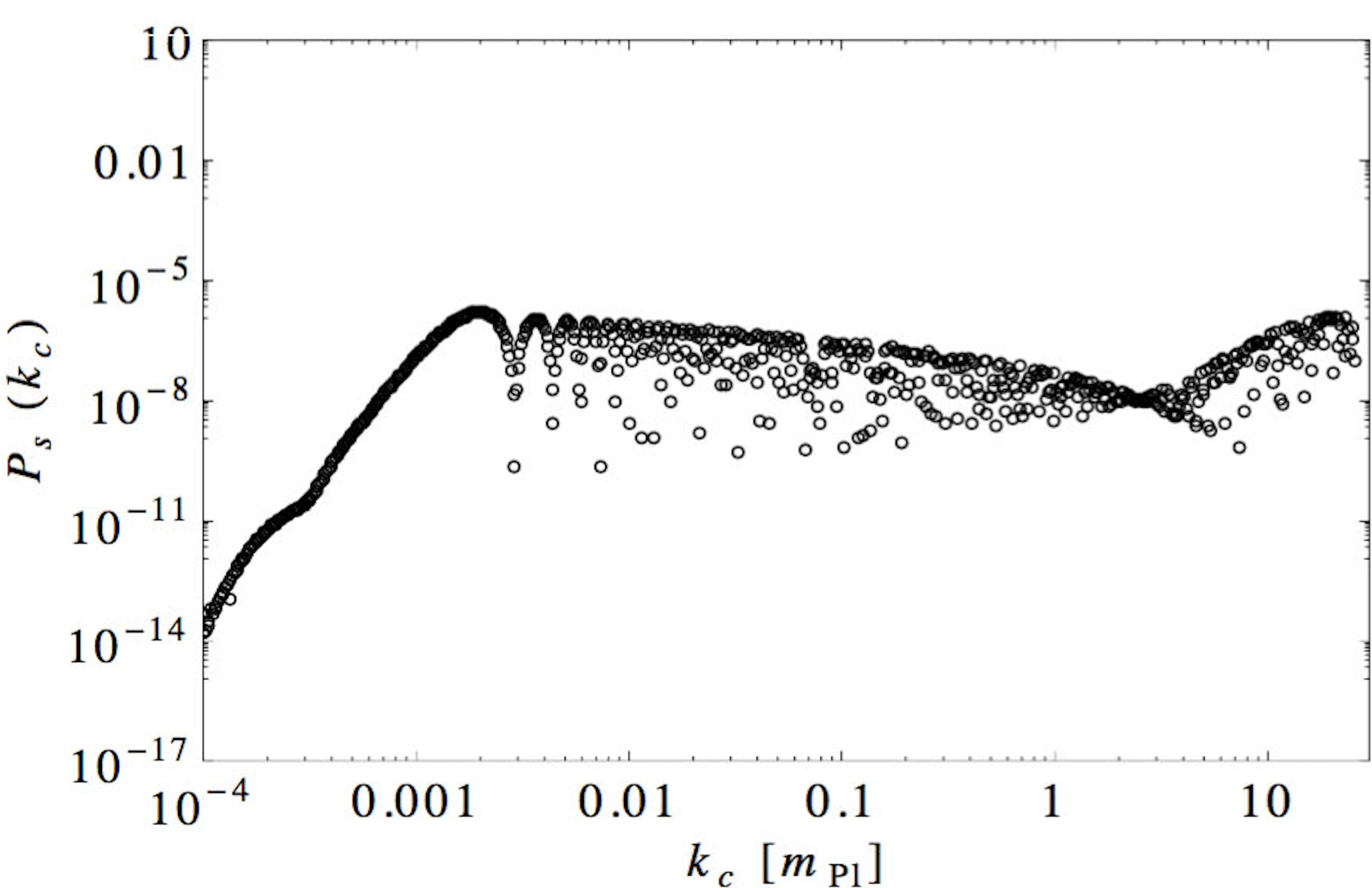}
\end{center}
\caption{Primordial scalar power spectrum in the deformed algebra approach with the Unruh-like MDR $~~$ $\mathcal{F}(k_{\varphi}) = k_{0} \tanh\left[\left(k_{c}/(a(t)k_{0})\right)^{p}\right]^{1/p}$, $k_{0}=10$ and $p=1$.}
\label{scalar power spectrum}
\end{figure}

Modified relations need to be applied to physical quantities and are therefore introduced in the Mukhanov-Sasaki equation by the replacement: $k_{\varphi}\rightarrow \mathcal{F}(k_{\varphi})$. The $\mathcal{F}(k_{\varphi})$ function depends on the considered dispersion relation, the standard case corresponding to $\mathcal{F}(k_{\varphi})=k_{\varphi}$. The Mukhanov-Sasaki equation in the Deformed Algebra approach then becomes:

\begin{equation}
v_{k}''(\eta) + \left(\Omega(\eta) a^{2}(\eta) \mathcal{F}(k_{\varphi})^{2} - \dfrac{z_{T/S}''(\eta)}{z_{T/S}(\eta)} \right) v_{k}(\eta) =0,
\end{equation}

where $\Omega(\eta) = 1 - 2 \rho(\eta) / \rho_{c}$. It is usually thought that the Deformed Algebra approach is excluded by data because of an exponential increase of the power spectra in the ultraviolet regime due to the $\Omega$ factor. One should however keep in mind that what is excluded is just a very specific scenario in which anisotropies, backreaction, and more importantly the trans-planckian problem, have been neglected. Namely we have shown that considering the MDR

\begin{equation}
\mathcal{F}(k_{\varphi}) = k_{0} \tanh\left[\left(\dfrac{k_{\varphi}}{k_{0}}\right)^{p}\right]^{\dfrac{1}{p}} = k_{0} \tanh\left[\left(\dfrac{k_{c}}{a(t)k_{0}}\right)^{p}\right]^{\dfrac{1}{p}},
\end{equation}

where $k_{0}$ is the transition scale in physical coordinates and $p$ determines the sharpness of the transition, cures the pathological behaviour of the deformed algebra, and leads to almost scale-invariant spectra in the observed regime both for tensor and scalar perturbations, as it can be seen in Figs. \ref{tensor power spectrum} and \ref{scalar power spectrum}.

This is a significant result in the sense that the Deformed Algebra approach is now shown, when trans-planckian effects are taken into account, to be possibly in agreement with data. This also shows that the spectra are actually highly sensitive to quantum gravity effects and a better understanding of the trans-planckian behaviour from the full LQG theory is necessary to go ahead in this direction.

\section*{Acknowledgments}

K. Martineau is supported by a grant from the C.F.M fundation.


\end{document}